\documentclass[journal]{IEEEtran}

\usepackage[utf8]{inputenc}
\usepackage[T1]{fontenc}
\usepackage{amsmath,amssymb,amsfonts}
\usepackage{amsfonts}
\usepackage{amssymb}
\usepackage{graphicx}
\usepackage[noadjust]{cite}
\usepackage{times}
\usepackage{bm}
\usepackage{algorithm}
\usepackage{algpseudocode}
\usepackage{mathtools}

\usepackage{subcaption}

\usepackage{tikz}
\usepackage{pgfplots}
\pgfplotsset{compat=newest}

\usetikzlibrary{spy}

\def\plotwidth{.5\textwidth}
\def\plotheight{.45\textwidth}

\def\plotsize{0.56\linewidth}

\makeatletter
\newcommand*\rel@kern[1]{\kern#1\dimexpr\macc@kerna}
\newcommand*\widebar[1]{%
	\begingroup
	\def\mathaccent##1##2{%
		\rel@kern{0.8}%
		\overline{\rel@kern{-0.5}\macc@nucleus\rel@kern{0.2}}%
		\rel@kern{-0.2}%
	}%
	\macc@depth\@ne
	\let\math@bgroup\@empty \let\math@egroup\macc@set@skewchar
	\mathsurround\z@ \frozen@everymath{\mathgroup\macc@group\relax}%
	\macc@set@skewchar\relax
	\let\mathaccentV\macc@nested@a
	\macc@nested@a\relax111{#1}%
	\endgroup
}
\makeatother

\newcommand\eqa{\stackrel{\mathclap{\normalfont\mbox{\tiny (a)}}}{=}}

\usepackage{glossaries}
\newacronym{ris}{RIS}{reconfigurable intelligent surface}

\def\ka#1{\textcolor{black}{#1}}
\def\kr#1{\textcolor{black}{#1}}
\def\krb#1{\textcolor{black}{#1}}

\begin{document}

 \title{{TRICE: A Channel Estimation Framework for RIS-Aided Millimeter-Wave MIMO Systems}}

\author{Khaled Ardah, Sepideh Gherekhloo, André L. F. de Almeida, and Martin Haardt
	
	\thanks{{The authors gratefully acknowledge the support of the German Research Foundation (DFG) under contract no.~HA 2239/6-2 (EXPRESS II) and the support of CAPES/PRINT (Grant no. 88887.311965/2018-00). The research of André L. F. de Almeida is partially supported by the CNPq (Grant no. 306616/2016-5).  }}
	\thanks{K. Ardah, S. Gherekhloo, and M. Haardt are with Communications Research Laboratory (CRL), TU Ilmenau, Ilmenau, Germany (e-mail: \{khaled.ardah, sepideh.gherekhloo, martin.haardt\}@tu-ilmenau.de). A.  de Almeida is with Wireless Telecom Research Group (GTEL), Federal University of Cear\'a, Fortaleza, Brazil (e-mail:  andre@gtel.ufc.br).}}

\maketitle

\begin{abstract}
\ka{We consider the channel estimation problem in point-to-point \gls{ris}-aided millimeter-wave (mmWave) MIMO systems. By exploiting the low-rank nature of mmWave channels in the angular domains, we propose a non-iterative Two-stage RIS-aided Channel Estimation (TRICE) framework, where every stage is formulated as a multidimensional direction-of-arrival (DOA) estimation problem. As a result, our TRICE framework is very general in the sense that any efficient multidimensional DOA estimation solution can be readily used in every stage to estimate the associated channel parameters. Numerical results show that the TRICE framework has a lower training overhead and a lower computational complexity, as compared to benchmark solutions.} 
\end{abstract}	

	
%\vspace{-5pt}
\section{Introduction}
Reconfigurable intelligent surfaces (RISs) have been proposed recently as {a} cost-effective {technology} for reconfiguring the wireless propagation channel between transceivers \cite{comMagazine,huang2019holographic,irs,Emil,access,Survey,wu2020intelligent}. 
\ka{In \gls{ris}-aided systems, an accurate channel state information (CSI) is required at transceivers to enable efficient signal processing techniques, e.g., beamforming and resource allocation.} However, the acquisition of CSI in such systems faces several challenges. For instance, assuming a passive RIS implementation, to reduce the RIS cost and complexity, the propagation channel can only be sensed and estimated at the receiver. Furthermore, the large number of channel coefficients to be estimated limits the feasibility of CSI acquisition within a practical coherence time, since an RIS is expected to have a massive number of passive reflecting elements. Recently, channel estimation methods for \gls{ris}-aided systems have been proposed, e.g., using least-squares (LS) based methods as {in} \cite{LSOnOff,MVDR,OFDM,andre}, or minimum mean squared error based methods as in \cite{MMSE}. However, these works require the number of training subframes to be, at least, equal to the number of \glspl{ris} reflecting elements, which is a limiting factor in practice. 

\ka{In millimeter-wave (mmWave) communications \cite{heath_overview,Rappaport,ardah_tvt,sepideh,jianshu,ardah_unify}, it was observed that the MIMO propagation channel has a low-rank structure, due to the small number of scatterers. Such a low-rank structure can be exploited to reduce the channel training overhead and complexity, as it has been shown in \cite{deeplearning,OFDMCS,he2020channel,MatrixCom,MUUL,CSGrid}. In these works, every channel matrix is modeled as a summation of $L$ paths, where $L$ is much smaller than the number of transmit and receiver antennas, and every path is completely characterized by a direction-of-departure (DOD), a direction-of-arrival (DOA), and a complex path gain. Therefore, the channel estimation is formulated as a sparse recovery problem, for which compressed sensing (CS) techniques \cite{CS} can be used to efficiently recover the channel parameters \kr{using} a \kr{small} training overhead. In \cite{deeplearning}, the above problem is facilitated by assuming that the RIS has a few active elements, which, however, increases the deployment cost and the energy consumption of RIS-aided systems. In \cite{OFDMCS}, the authors assumed that the base station (BS)-to-RIS channel is perfectly known, while in \cite{he2020channel}, the cascade channel matrix is assumed to have a single path, i.e., $L = 1$. Differently, the authors in \cite{CSGrid} proposed a general sparse recovery formulation for $L \geq 1$ scenarios. In most of these works, however, the channel parameters are assumed to fall perfectly on a grid, which may never be true in practice. Therefore, there exists a trade-off between the estimation accuracy and the complexity, where both increase \kr{as a function of the} grid resolution. Due to the multidimensionality of the \kr{cascaded} channel, a 4D sensing matrix is required by the method proposed in \cite{CSGrid}, which makes it computationally prohibitive even with low grid resolutions.}

In this paper, we consider the channel estimation problem in a single-user \gls{ris}-aided mmWave MIMO communication system, similarly to \cite{CSGrid}, where the \gls{ris} has passive reflecting elements and the direct link between the BS and the mobile station (MS) is assumed to be blocked or pre-estimated by turning the \gls{ris} elements {off}, as in \cite{LSOnOff}. 
Using a structured channel training procedure, we propose a \textbf{T}wo-Stage \textbf{RI}S-aided \textbf{C}hannel \textbf{E}stimation (TRICE) framework for single-user mmWave MIMO communication systems. In the first stage, the DODs of the BS-to-\gls{ris} channel and {the} DOAs of the \gls{ris}-to-MS channel are first estimated. In the second stage, {by using} the estimated channel parameters in the first stage, \krb{the effective azimuth and elevation angles of the cascaded BS-to-RIS-to-MS channel at the RIS are estimated, one-by-one, including the effective complex path gains}. In both stages, we show that the parameter estimation can be {carried out \textit{via}} a {multidimensional DOA} estimation scheme, for which several solutions exist as in \cite{StESBRIT,UESBRIT,L21Norm,ardah_icassp19,NOMP,URA}, among many others. Detailed simulation results are provided, showing that the proposed {TRICE} framework has a lower training overhead and a lower computational complexity, as compared to benchmark methods.        

\section{System and Channel Models}

\begin{figure}
	\centering
	\includegraphics[width=1\linewidth]{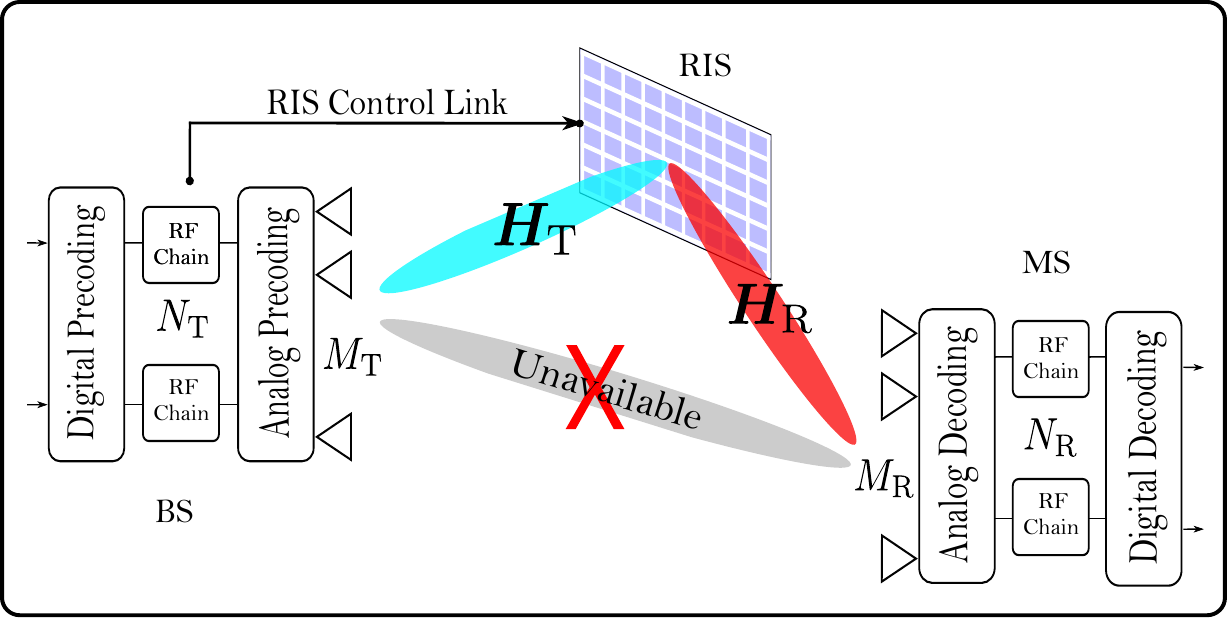}
	\caption{An \gls{ris}-aided mmWave MIMO communication system.}
	\label{fig:fig1}
	\vspace{-10pt}
\end{figure}

In this paper\footnote{\textbf{Notation.} Matrices (vectors) are represented by boldface capital (lowercase) letters, $\bm{A}^{T}$, $\bm{A}^{+}$, $\otimes$, $\diamond$, and $\odot$ denote the transpose, the Moore-Penrose pseudo-inverse, the Kronecker, the Khatri-Rao, and the Hadamard products, respectively, $\text{diag}\{\bm{a}\}$ forms a matrix by placing $\bm{a}$ {on} its main diagonal, and $\text{vec}\{\bm{A}\}$ vectorizes $\bm{A}$ by arranging its columns {on top of} each other. We define $[\bm{a}]_{[n]}$ as the $n$th entry of vector $\bm{a}$, $\bm{1}_N$ as the all ones vector of length $N$, $\bm{I}_N$ as the $N\times N$ identity matrix, $\mathcal{CN}(\bm{0},\bm{R})$ as the circularly symmetric complex Gaussian distribution with zeros mean and covariance matrix $\bm{R}$, and $\mathcal{U}(a_1,a_2)$ as the uniform distribution within the interval $[a_1,a_2]$. {Moreover}, the following properties are used: {Property 1}: $\text{vec}\{ \bm{A}\bm{B}\bm{C} \} = (\bm{C}^{T}\otimes\bm{A}) \text{vec}\{\bm{B}\}$. {Property 2}: $(\bm{A}\bm{B}\diamond \bm{C}\bm{D}) = (\bm{A}\otimes \bm{C})(\bm{B}\diamond \bm{D})$. {Property 3}: $(\bm{A}\otimes \bm{C})(\bm{B}\otimes \bm{D}) = (\bm{A}\bm{B}\otimes \bm{C}\bm{D})$.}, we consider a single-user mmWave MIMO communication system as depicted in Fig. \ref{fig:fig1}, where a BS equipped with $M_{\text{T}}$ antennas and $N_{\text{T}} \leq M_{\text{T}}$ RF chains is communicating with a MS that has $M_{\text{R}}$ antennas and $N_{\text{R}} \leq M_{\text{R}}$ RF chains. We assume that the direct link between the BS and the MS is unavailable (e.g., due to blockage) and the indirect link is {aided} by an \gls{ris} composed by $M_{\text{S}}$ phase shifters, which are arranged {uniformly} {on} a rectangular surface with $M^{\text{v}}_\text{S}$ vertical and $M^{\text{h}}_\text{S}$ horizontal elements such that $M_{\text{S}} = M^{\text{v}}_\text{S} \cdot M^{\text{h}}_\text{S}$.

We assume that the BS and the MS employ uniform linear arrays (ULAs)\footnote{The extension of the proposed {TRICE} framework to scenarios where {the} BS and/or {the} MS are {equipped} with URAs is straightforward.}. Let $\bm{H}_{\text{T}} \in \mathbb{C}^{M_{\text{S}}\times M_{\text{T}}}$ ($\bm{H}_{\text{R}} \in \mathbb{C}^{M_{\text{R}}\times M_{\text{S}}}$) denotes the mmWave MIMO channel between the BS (RIS) and the \gls{ris} (MS). \ka{Similarly to \cite{CSGrid}, $\bm{H}_{\text{T}}$ and $\bm{H}_{\text{R}}$ are modeled {according to the classical} Saleh-Valenzuela model \cite{SV} as
\begin{equation}\label{channels}
	\begin{aligned}
		\bm{H}_{\text{T}} &=   \sum_{\ell = 1}^{L_{\text{T}}} \alpha_{\text{T},\ell}  \bm{v}_{{\text{2D}}}(\mu^{\text{v}}_{\text{T},\ell}, \mu^{\text{h}}_{\text{T},\ell})  \bm{v}_{{\text{1D}}}(\psi_{\text{T},\ell})^{T} = \bm{B}_{\text{T}} \bm{G}_{\text{T}}\bm{A}^{T}_{\text{T}} \\
		\bm{H}_{\text{R}} &=  \sum_{\ell = 1}^{L_{\text{R}}} \alpha_{\text{R},\ell} \bm{v}_{{\text{1D}}}(\psi_{\text{R},\ell}) \bm{v}_{{\text{2D}}}(\mu^{\text{v}}_{\text{R},\ell}, \mu^{\text{h}}_{\text{R},\ell})^{T}  = \bm{A}_{\text{R}}  \bm{G}_{\text{R}} \bm{B}^{T}_{\text{R}}, 
	\end{aligned}
\end{equation}
where $\alpha_{\text{T},\ell}$ and $\alpha_{\text{R},\ell}$ are the complex path gains, $\psi_{\text{T},\ell}$ ($\psi_{\text{R},\ell}$) is the $\ell$th path DOD (DOA) spatial frequency at \kr{the} BS (MS), while $\mu^{\text{h}}_{\text{T},\ell} $ and $ \mu^{\text{v}}_{\text{T},\ell}$ ($\mu^{\text{h}}_{\text{R},\ell} $ and $ \mu^{\text{v}}_{\text{R},\ell}$) are the $\ell$th path azimuth and elevation DOAs (DODs) spatial frequencies at \kr{the} RIS\footnote{\ka{Let $d$ denotes the antenna spacing and $\lambda$ be the signal wavelength. Then, the spatial frequencies are defined as $\psi_{\text{X},\ell} = 2\pi\frac{d}{\lambda} \cos(\phi_{\text{X},\ell})$, $\mu^{\text{h}}_{\text{X},\ell} = 2\pi\frac{d}{\lambda} \cos(\theta^{\text{h}}_{\text{X},\ell})$, and $\mu^{\text{v}}_{\text{X},\ell} = 2\pi\frac{d}{\lambda} \sin(\theta^{\text{h}}_{\text{X},\ell}) \cos(\theta^{\text{v}}_{\text{X},\ell})$, where $\phi_{\text{X},\ell} \in [-180^{\circ},180^{\circ}]$ is the $\ell$th path angle in the angular domain, while $\theta^{\text{h}}_{\text{X},\ell} \in [-180^{\circ},180^{\circ}]$ ($\theta^{\text{v}}_{\text{X},\ell} \in [-90^{\circ},90^{\circ}]$) is the $\ell$th path azimuth (elevation) angle at \kr{the} RIS in the angular domain.}}. Moreover, $\bm{v}_{{\text{2D}}}(\mu^{\text{v}}_{\text{X},\ell},\mu^{\text{h}}_{\text{X},\ell}) = \bm{v}_{{\text{1D}}}(\mu^{\text{v}}_{\text{X},\ell}) \diamond \bm{v}_{{\text{1D}}}(\mu^{\text{h}}_{\text{X},\ell}) \in\mathbb{C}^{M_{\text{S}}}$ and $\bm{v}_{{\text{1D}}}(\psi_{\text{X},\ell})  \in\mathbb{C}^{M_{\text{X}}}$ are the functions representing the 2D and the 1D array steering vectors, respectively, where $\text{X} \in \{\text{T},\text{R}\}$. For a given spatial frequency $\nu$, the steering vector $\bm{v}_{{\text{1D}}}(\nu)$ is given as $\bm{v}_{{\text{1D}}}(\nu) = [1, e^{j \nu}, \dots,e^{j (M-1)  \nu}]^{T} \in\mathbb{C}^{M}.$
In (\ref{channels}), $\bm{H}_{\text{T}}$ and $\bm{H}_{\text{R}}$ are written in a compact form by letting $\bm{A}_{\text{X}} = [\bm{v}_{{\text{1D}}}({\psi}_{\text{X},1}),\dots,\bm{v}_{{\text{1D}}}({\psi}_{\text{X},L_{\text{X}}})] \in\mathbb{C}^{M_{\text{X}}\times L_{\text{X}}}$, $\bm{B}_{\text{X}} = \bm{B}^{\text{v}}_{\text{X}} \diamond \bm{B}^{\text{h}}_{\text{X}} \in\mathbb{C}^{M_{\text{S}}\times L_{\text{X}}}$, $\bm{B}^{\text{y}}_{\text{X}} = [\bm{v}_{{\text{1D}}}({\mu}^{\text{y}}_{\text{X},1}),\dots,\bm{v}_{{\text{1D}}}({\mu}^{\text{y}}_{\text{X},L_{\text{X}}})] \in\mathbb{C}^{M^{\text{y}}_{\text{S}}\times L_{\text{X}}}$, and $\bm{G}_{\text{X}} = \text{diag}\{ {\alpha}_{\text{X},1},\dots, {\alpha}_{\text{X},L_{\text{X}}}  \}$, where $\text{y}\in\{\text{v},\text{h}\}$. }

We assume a block-fading channel, where $\bm{H}_{\text{T}}$ and $\bm{H}_{\text{R}}$ remain constant during each block and change from block to block. To estimate $\bm{H}_{\text{T}}$ and $\bm{H}_{\text{R}}$, we conduct a channel training procedure at the beginning of each block, which {comprises} $K$ frames divided into $K_{\text{T}}\cdot K_{\text{S}}$ subframes, i.e., $K = K_{\text{S}}\cdot K_{\text{T}}$. At the BS, we assume that {a single} RF chain is used during the channel training procedure, to reduce the energy consumption, which implies that a single training vector is transmitted in every subframe. Let $ \widetilde{\bm{F}} = [\widetilde{\bm{f}}_1,\dots,\widetilde{\bm{f}}_{K_{T}} ] \in \mathbb{C}^{M_{\text{T}}\times K_{\text{T}}}$ be the matrix holding the $K_{\text{T}}$ analog training vectors of the BS, with $\big|[\widetilde{\bm{f}}_{t}]_{[i]}\big| = \frac{1}{\sqrt{M_{\text{T}}}},\forall t, i$, {and} $\widetilde{\bm{F}}^{H} \widetilde{\bm{F}} = \bm{I}_{K_{\text{T}}}$. Moreover, let $\bm{Q} = [\bm{q}_1 ,\dots,\bm{q}_{K_{\text{S}}}] \in \mathbb{C}^{M_{\text{S}} \times K_{\text{S}}} $ be the matrix holding the $K_{\text{S}}$ {phase shift} vectors of the \gls{ris}, with $\big|[{\bm{q}}_{s}]_{[j]}\big| = \frac{1}{\sqrt{M_{\text{S}}}}, \forall s, j$. We propose to design $\bm{Q}$ to have a Kronecker structure as 
\begin{align}\label{Q}
\bm{Q} = \bm{Q}_{\text{v}} \otimes \bm{Q}_{\text{h}}  \in \mathbb{C}^{M_{\text{S}} \times K_{\text{S}}},
\end{align}  
where $\bm{Q}_{\text{v}} \in \mathbb{C}^{M^{\text{v}}_{\text{S}} \times K^{\text{v}}_{\text{S}}}$, $\bm{Q}_{\text{h}} \in \mathbb{C}^{M^{\text{h}}_{\text{S}} \times K^{\text{h}}_{\text{S}}}$, and $K_{\text{S}} = K^{\text{v}}_{\text{S}}\cdot K^{\text{h}}_{\text{S}}$. Such a design structure will be exploited in Section \ref{Sec:proposed} to obtain a low-complexity channel estimation method.

The received signal at the MS at the $(s,t)$th subframe, $s\in\{1,\dots,K_{\text{S}}\}$, $t\in\{1,\dots,K_{\text{T}}\}$, is given as
\begin{align}
\bm{y}_{s,t} & = \bm{W}^{T} \bm{H}_{\text{R}} \text{diag}\{\bm{q}_s\} \bm{H}_{\text{T}} \widetilde{\bm{f}}_{t} p_{t} + \bm{z}_{s,t} \in \mathbb{C}^{N_{\text{R}}},
\end{align}
where $\bm{W} \in \mathbb{C}^{M_{\text{R}}\times N_{\text{R}}}$ is the decoding matrix, $p_{t} \in \mathbb{C}$ is the unit-power pilot signal, and $\bm{z}_{s,t} \in \mathbb{C}^{N_{\text{R}}}$ is the additive white Gaussian noise vector {having zero-mean circularly symmetric complex-valued entries} with variance $\sigma^2$. Let $\bm{F} = [\widetilde{\bm{f}}_1 {p}_1,\dots,\widetilde{\bm{f}}_{K_{T}} {p}_{K_{\text{T}}}]$. Then, by stacking $\bm{y}_{s,t}, \forall t$, {on top of} each other as $\bm{y}_{s} =  [\bm{y}^{T}_{s,1},\dots,\bm{y}^{T}_{s,K_{\text{T}}}]^{T}$, we have 
\begin{align}\label{ys}
\bm{y}_s  = (\bm{F}^{T} \bm{H}^{T}_{\text{T}} \diamond \bm{W}^{T} \bm{H}_{\text{R}} ) \bm{q}_{s} + \bm{z}_{s} \in \mathbb{C}^{N_{\text{R}}K_{\text{T}}},
\end{align}
where $\bm{z}_{s} =  [\bm{z}^{T}_{s,1},\dots,\bm{z}^{T}_{s,K_{\text{T}}}]^{T} $. Let $\bm{H} = (\bm{H}^{T}_{\text{T}} \diamond \bm{H}_{\text{R}}) \in \mathbb{C}^{M_{\text{R}}M_{\text{T}}\times M_{\text{S}}}$ and $\bm{Z} = [\bm{z}_1,\dots,\bm{z}_{K_{\text{S}}}]$. Then, by stacking $\bm{y}_s, \forall s$, as $\bm{Y} = [\bm{y}_1,\dots, \bm{y}_{K_{\text{S}}}]$ and applying Property 2, we have
\begin{align}\label{eq1}
\bm{Y} =  (\bm{F}^{T} \otimes \bm{W}^{T}) \bm{H}   \bm{Q} + \bm{Z} \in \mathbb{C}^{N_{\text{R}} K_{\text{T}} \times K_{\text{S}}}.
\end{align}

Our main goal is to estimate $\bm{H}$ from  (\ref{eq1}). One direct solution is to use the LS-based method. By applying Property 1, the vectorized form of (\ref{eq1}) can be written as $\bm{y} = \bm{\Upsilon}  \bm{h} + \bm{z}$, where $\bm{\Upsilon} = (\bm{Q}^{T} \otimes \bm{F}^{T} \otimes \bm{W}^{T}) \in \mathbb{C}^{N_{\text{R}}K_{\text{T}}K_{\text{S}} \times M_{\text{R}}M_{\text{T}} M_{\text{S}}}$, $\bm{h} = \text{vec}\{\bm{H}\}$, and $\bm{z} = \text{vec}\{\bm{Z}\}$. Therefore, an estimate to the channel vector $\bm{h}$ can be obtained as $\widehat{\bm{h}}_{\text{LS}} =  \bm{\Upsilon}^{+}\bm{y}$, which requires $K = K_{\text{T}}K_{\text{S}} \geq \frac{M_{\text{R}}M_{\text{T}} M_{\text{S}}}{N_{\text{R}}}$ to have an accurate channel estimate. Such an approach, however, becomes impractical in a massive MIMO setup, since it requires a large number of training subframes $K$ and a long channel coherence time.

\section{Proposed {TRICE} Framework}\label{Sec:proposed}
\ka{From (\ref{channels}), the \kr{cascaded} channel matrix $\bm{H}$ can be written as   
\begin{align}\label{h}
\bm{H} & =  (\bm{A}_{\text{T}} \bm{G}_{\text{T}}\bm{B}^{T}_{\text{T}} \diamond \bm{A}_{\text{R}} \bm{G}_{\text{R}}\bm{B}^{T}_{\text{R}})   \eqa  (\bm{A}_{\text{T}} \otimes \bm{A}_{\text{R}}) \bm{G} \bm{B},
\end{align}
where $\bm{G} = (\bm{G}_{\text{T}} \otimes \bm{G}_{\text{R}}) \in \mathbb{C}^{L \times L} $, $\bm{B} = (\bm{B}^{T}_{\text{T}} \diamond \bm{B}^{T}_{\text{R}} ) \in \mathbb{C}^{L\times M_{\text{S}}} $, $L = L_{\text{R}} {L}_{\text{T}}$, and $\eqa$ is obtained from Property 2. {Using (\ref{h}) and applying} Property~3, we have 
\begin{align}\label{eq2}
\bm{Y} = \bm{A} \bm{X} + \bm{Z} \in \mathbb{C}^{N_{\text{R}} K_{\text{T}} \times K_{\text{S}}}, 
\end{align}
where $\bm{A} = (\bm{F}^{T}\bm{A}_{\text{T}} \otimes \bm{W}^{T} \bm{A}_{\text{R}}) $ and $\bm{X} = \bm{G} \bm{B}  \bm{Q}$. Observing (\ref{eq2}), we can see that $\bm{A}$ is completely characterized by the frequency vectors defined as ${\bm{\psi}}_{\text{T}} = [{\psi}_{\text{T},1}, \dots, {\psi}_{\text{T},L_{\text{T}}}]^{T}$ and ${\bm{\psi}}_{\text{R}} = [{\psi}_{\text{R},1}, \dots, {\psi}_{\text{R},L_{\text{R}}}]^{T}$. Therefore, estimating ${\bm{\psi}}_{\text{T}}$ and ${\bm{\psi}}_{\text{R}}$ from (\ref{eq2}) is, in fact, a {2D DOA} estimation problem, where several methods exist in the literature, such as in \cite{StESBRIT,UESBRIT,L21Norm,ardah_icassp19,NOMP,URA}, among many others. 
For instance, the DFT-beamspace ESPRIT methods of \cite{StESBRIT,UESBRIT} can be readily applied to estimate  ${\bm{\psi}}_{\text{T}}$ and ${\bm{\psi}}_{\text{R}}$ in a closed form with guaranteed automatic pairing \cite{HaardtSSD}. While subspace-based methods perform asymptotically optimal, they suffer from a performance degradation in the case of difficult scenarios such as high noise power and \kr{small} number of measurement vectors. Alternatively, CS techniques \cite{L21Norm,ardah_icassp19,NOMP,URA} have been shown to provide an attractive alternative to subspace-based methods, \kr{yielding} good estimation performance even in difficult scenarios. To show this, we note that (\ref{eq2}) can be written in a sparse form as  
\begin{align}\label{sparse}
	\bm{Y} \approx  (\bm{F}^{T} \widebar{\bm{A}}_{\text{T}} \otimes \bm{W}^{T} \widebar{\bm{A}}_{\text{R}}) \text{ } \widebar{\bm{X}}  + \bm{Z} \in \mathbb{C}^{N_{\text{R}} K_{\text{T}} \times K_{\text{S}}},
\end{align}
where $\widebar{\bm{A}}_{\text{T}} \in \mathbb{C}^{M_{\text{T}} \times \bar{L}_{\text{T}}} $ and $\widebar{\bm{A}}_{\text{R}} \in \mathbb{C}^{M_{\text{R}} \times \bar{L}_{\text{R}}}$ represent two dictionary matrices, in which $\bar{L}_{\text{T}} \gg {L}_{\text{T}}$ and $\bar{L}_{\text{R}} \gg {L}_{\text{R}}$ define the number of grid points or, in other words, the grid resolution, while $\widebar{\bm{X}} \in \mathbb{C}^{\bar{L}_{\text{T}} \bar{L}_{\text{R}} \times K_{\text{S}}}$ is an ${L}$ row-sparse matrix \cite{L21Norm}. Here, (\ref{sparse}) can be written with equality if, and only if, the true angles ${\bm{\psi}}_{\text{T}}$ and ${\bm{\psi}}_{\text{R}}$ fall perfectly on the grid points. In this latter case, the $k$th nonzero row of $\widebar{\bm{X}}$ equals to the $k$th row of $\bm{X} $. Note that (\ref{sparse}) \kr{corresponds to} a sparse recovery problem. Therefore, known CS techniques, e.g., \cite{L21Norm,ardah_icassp19,NOMP,URA}, including the OMP method \cite{OMPComp} can readily be applied to estimate $\widebar{\bm{X}}$, as well as, ${\bm{\psi}}_{\text{T}}$ and ${\bm{\psi}}_{\text{R}}$, with automatic pairing. Since $L \ll M_{\text{R}} M_{\text{T}}$, due to the low-rank nature of the mmWave channels, only a few measurements (training overhead) are required, i.e., $N_{\text{R}} K_{\text{T}} \approx \mathcal{O}(L\log(\bar{L}_\text{R} \bar{L}_\text{T} /L)) \ll M_{\text{R}} M_{\text{T}}$ \cite{CSBounds}\footnote{Note that the recoverability guarantee of $\widebar{\bm{X}}$ in (\ref{sparse}) can be improved by probably designing the sensing matrix $\widebar{\bm{A}} = (\bm{F}^{T} \widebar{\bm{A}}_{\text{T}} \otimes \bm{W}^{T} \widebar{\bm{A}}_{\text{R}})$, as in \cite{ardah_icassp2020,CSSMCM}, which is out of the scope of this paper.}.}

\ka{To proceed, let ${\widehat{\bm{\psi}}}_{\text{T}}$ and $\widehat{\bm{\psi}}_{\text{R}}$ denote the estimated frequency vectors of ${{\bm{\psi}}}_{\text{T}}$ and ${\bm{\psi}}_{\text{R}}$. Then, we construct $\widehat{\bm{A}}_{\text{T}}$, $\widehat{\bm{A}}_{\text{R}}$, and $\widehat{\bm{A}} = (\bm{F}^{T} \widehat{\bm{A}}_{\text{T}} \otimes \bm{W}^{T} \widehat{\bm{A}}_{\text{R}})$. Therefore, to estimate $\bm{H}$ in (\ref{h}), an estimate of $\bm{G}$ and $\bm{B}$ is required. Let us assume that ${\bm{\psi}}_{\text{T}}$ and ${\bm{\psi}}_{\text{R}}$ are estimated perfectly and that the $\text{rank}\{\widehat{\bm{A}}\} \geq L$. Then, multiplying (\ref{eq2}) by $\widehat{\bm{A}}^{+}$ from the left-hand-side we get } 
\begin{align}\label{eq4}
 \underline{\bm{Y}} =  \widehat{\bm{A}}^{+} \bm{Y} =  \bm{G} \bm{B}  \bm{Q} +  \underline{\bm{Z}} \in \mathbb{C}^{L \times K_{\text{S}} },
\end{align} 
where $\underline{\bm{Z}} = \widehat{\bm{A}}^{+} \bm{Z} \in \mathbb{C}^{L \times K_{\text{S}}} $ is the filtered noise. 
\ka{Since $\bm{G} = \bm{G}^{T}$, due to its diagonal structure, we can write $\underline{\bm{Y}}^{T} $ as }
\begin{align}\label{Bunder}
\underline{\bm{Y}}^{T} = \bm{Q}^{T}  {\bm{B}}^{T} \bm{G}   + \underline{\bm{Z}}^{T} \in \mathbb{C}^{K_\text{S} \times L}.
\end{align}

\ka{Note that, ${\bm{B}}^{T} \in \mathbb{C}^{M_{\text{S}}\times L} $ can be written as}
\begingroup\makeatletter\def\f@size{8.5}\check@mathfonts
\def\maketag@@@#1{\hbox{\m@th\small\normalfont#1}}% 
\begin{align}
{\bm{B}}^{T} & = \big[ (\bm{b}^{T}_{\text{T},1} \diamond \bm{b}^{T}_{\text{R},1})^{T}, \dots,  (\bm{b}^{T}_{\text{T},1} \diamond \bm{b}^{T}_{\text{R},L_\text{R}})^{T}, \dots, (\bm{b}^{T}_{\text{T},L_\text{T}} \diamond \bm{b}^{T}_{\text{R},L_\text{R}} )^{T} \big]  \nonumber \\ 
& = \big[ 
(\bm{b}_{\text{T},1} \odot \bm{b}_{\text{R},1}),
\dots,  
(\bm{b}_{\text{T},1} \odot \bm{b}_{\text{R},L_\text{R}}), 
\dots,   
(\bm{b}_{\text{T},L_\text{T}} \odot \bm{b}_{\text{R},L_\text{R}})
\big], 
\end{align}\endgroup
\ka{where $\bm{b}_{\text{T},\ell} =  \bm{v}_{\text{1D} }({\mu}^{\text{v}}_{\text{T},\ell} ) \diamond \bm{v}_{\text{1D} }({\mu}^{\text{h}}_{\text{T},\ell} ) $ and $\bm{b}_{\text{R},k} =  \bm{v}_{\text{1D} }({\mu}^{\text{v}}_{\text{R},k} ) \diamond \bm{v}_{\text{1D} }({\mu}^{\text{h}}_{\text{R},k} )$ are the $\ell$th and the $k$th {column} vectors of $\bm{B}_\text{T}$ and $\bm{B}_\text{R}$, respectively, $\ell \in \{1,\dots, L_\text{T} \}$, $k \in \{1,\dots, L_\text{R} \}$, i.e., }
\begingroup\makeatletter\def\f@size{8}\check@mathfonts
\def\maketag@@@#1{\hbox{\m@th\small\normalfont#1}}% 
\begin{align*}
	\bm{b}_{\text{T},\ell} = \begin{bmatrix}
		1 \cdot\bm{v}_{\text{1D} }({\mu}^{\text{h}}_{\text{T},\ell} ) \\
		e^{j {\mu}^{\text{v}}_{\text{T},\ell} } \cdot \bm{v}_{\text{1D} }({\mu}^{\text{h}}_{\text{T},\ell} ) \\
		\vdots  \\
		e^{j  (M^{\text{v}}_\text{S} - 1) {\mu}^{\text{v}}_{\text{T},\ell}  } \cdot \bm{v}_{\text{1D} }({\mu}^{\text{h}}_{\text{T},\ell} ) 
	\end{bmatrix}, \bm{b}_{\text{R},k} = \begin{bmatrix}
		1  \cdot\bm{v}_{\text{1D} }({\mu}^{\text{h}}_{\text{R},k} ) \\
		e^{j {\mu}^{\text{v}}_{\text{R},k} } \cdot \bm{v}_{\text{1D} }({\mu}^{\text{h}}_{\text{R},k} ) \\
		\vdots  \\
		e^{j  (M^{\text{v}}_\text{S} - 1) {\mu}^{\text{v}}_{\text{R},k}  } \cdot \bm{v}_{\text{1D} }({\mu}^{\text{h}}_{\text{R},k} ) 
	\end{bmatrix} .  
\end{align*}\endgroup

\ka{Therefore, the $n$th column of ${\bm{B}}^{T}$, i.e., $ {\bm{b}}_n = ( \bm{b}_{\text{T},\ell} \odot  \bm{b}_{\text{R},k} )$ has a Khatri-Rao structure given as }
\begin{align*}
	{{\bm{b}}}_n & = \begin{bmatrix}
		1  & \cdot  &  \bm{v}_{\text{1D} }({\mu}^{\text{h}}_{\text{T},\ell} + {\mu}^{\text{h}}_{\text{R},k} )  \\
		e^{j ({\mu}^{\text{v}}_{\text{T},\ell} + {\mu}^{\text{v}}_{\text{R},k}) }   &\cdot&  \bm{v}_{\text{1D} }({\mu}^{\text{h}}_{\text{T},\ell} + {\mu}^{\text{h}}_{\text{R},k} )   \\
		& \vdots & \\
		e^{j  (M^{\text{v}}_\text{S} - 1) ({\mu}^{\text{v}}_{\text{T},\ell} + {\mu}^{\text{v}}_{\text{R},k})  }  &\cdot&  \bm{v}_{\text{1D} }({\mu}^{\text{h}}_{\text{T},\ell} + {\mu}^{\text{h}}_{\text{R},k} ) 
	\end{bmatrix},
\end{align*} 
\ka{where $n = (\ell-1)\cdot L_\text{R} + k \in \{1,\dots, L\}$. 
Let ${\mu}^{\text{v}}_{n} = {\mu}^{\text{v}}_{\text{T},\ell} + {\mu}^{\text{v}}_{\text{R},k}$ and ${\mu}^{\text{h}}_{n} = {\mu}^{\text{h}}_{\text{T},\ell} + {\mu}^{\text{h}}_{\text{R},k}$. Then, we have }
\begin{align*}
	{{\bm{b}}}_n & = \bm{v}_{\text{1D} }({\mu}^{\text{v}}_{n} ) \diamond \bm{v}_{\text{1D} }({\mu}^{\text{h}}_{n} )     \in \mathbb{C}^{M_\text{S}}, 
\end{align*} 
\ka{where $\bm{v}_{\text{1D} }({\mu}^{\text{v}}_{n} ) \in \mathbb{C}^{M^{\text{v}}_\text{S}}$ and $ \bm{v}_{\text{1D} }({\mu}^{\text{h}}_{n} ) \in \mathbb{C}^{M^{\text{h}}_\text{S}}$. Accordingly, ${\bm{B}}^{T} = ({\bm{B}}^{\text{v}} \diamond  {\bm{B}}^{\text{h}})$, in which ${\bm{B}}^{\text{v}} = [\bm{v}_{\text{1D} }({\mu}^{\text{v}}_{1} ), \dots, \bm{v}_{\text{1D} }({\mu}^{\text{v}}_{ L } ) ]$, ${\bm{B}}^{\text{h}} = [\bm{v}_{\text{1D} }({\mu}^{\text{h}}_{1} ), \dots, \bm{v}_{\text{1D} }({\mu}^{\text{h}}_{ L} ) ] $, and (\ref{Bunder}) can be written as }
\begin{align}\label{Y3new0}
\underline{\bm{Y}}^{T} & \eqa (\bm{Q}^{T}_{\text{v}} {\bm{B}}^{\text{v}} \diamond  \bm{Q}^{T}_{\text{h}} {\bm{B}}^{\text{h}}) \bm{G}   + \underline{\bm{Z}}^{T} \in \mathbb{C}^{K_\text{S} \times L},
\end{align}
\ka{where $\eqa$ is obtained by utilizing the structure of $\bm{Q}$ in (\ref{Q}) and Property~2. Similarly to (\ref{eq2}), the first term on the right-hand-side of (\ref{Y3new0}) is completely characterized by the frequency vectors $\bm{\mu}^{\text{v}} = [\mu^{\text{v}}_{1}, \dots, \mu^{\text{v}}_{L}]$ and  $\bm{\mu}^{\text{h}} = [\mu^{\text{h}}_{1}, \dots, \mu^{\text{h}}_{L}]$. Therefore, $\bm{\mu}^{\text{v}}$ and $\bm{\mu}^{\text{h}}$ can be estimated using the same methods discussed above. However, it should be noted that the joint estimation of $\bm{\mu}^{\text{v}}$ and $\bm{\mu}^{\text{h}}$ does not guarantee the automatic pairing with the pre-estimated frequency vectors ${\bm{\psi}}_{\text{T}}$ and ${\bm{\psi}}_{\text{R}}$. To overcome this issue, we utilize the diagonal structure of the $\bm{G}$ matrix in (\ref{Y3new0}) and propose to estimate  $\bm{\mu}^{\text{v}}$ and $\bm{\mu}^{\text{h}}$ sequentially, where the $n$th entries $\mu^{\text{v}}_{n}$ and $\mu^{\text{h}}_{n}$ can be jointly estimated from the $n$th \kr{column} vector of $\underline{\bm{Y}}^{T}$ in (\ref{Y3new0}), i.e., $\underline{\bm{y}}_{n}$ that is given as }
\begin{align}\label{y3n}
\underline{\bm{y}}_{n} & = ( \bm{Q}^{T}_{\text{v}} \bm{v}_{\text{1D} }({\mu}^{\text{v}}_{n} ) \diamond  \bm{Q}^{T}_{\text{h}} \bm{v}_{\text{1D} }({\mu}^{\text{h}}_{n} ) ) \alpha_{n}  + \underline{\bm{z}}_{n} \in \mathbb{C}^{K_\text{S}},
\end{align} 
\ka{where $\alpha_{n}$ is the $n$th diagonal entry of $\bm{G}$ and $\underline{\bm{z}}_{n}$ is the  $n$th \kr{column} vector of $\underline{\bm{Z}}^{T}$. Note that, due to the Kronecker structure of $\bm{Q}$ in (\ref{Q}), it is possible to apply the DFT-beamspace ESPRIT method of \cite{StESBRIT} on (\ref{y3n}) to \kr{obtain} closed form estimates of $\mu^{\text{v}}_{n}$ and $\mu^{\text{h}}_{n}$.
Next, for given $\hat{\mu}^{\text{h}}_{n}$ and $\hat{\mu}^{\text{v}}_{n}$, the $n$th path gain $\alpha_{n}$ can be estimated from (\ref{y3n}) using LS as }
\begin{align}\label{alphn}
\hat{\alpha}_{n} =  \big(  \bm{Q}^{T}_{\text{v}} \bm{v}_{\text{1D} }(\hat{\mu}^{\text{v}}_{n} ) \diamond  \bm{Q}^{T}_{\text{h}} \bm{v}_{\text{1D} }(\hat{\mu}^{\text{h}}_{n}  ) \big)^{+}   \underline{\bm{y}}_{n}.
\end{align}

\ka{Finally, the $\bm{B}$ matrix in (\ref{h}) can be reconstructed as $\widehat{\bm{B}} =  (\widehat{{\bm{B}}}^{\text{v}} \diamond \widehat{{\bm{B}}}^{\text{h}})^{T} \in \mathbb{C}^{L \times M_{\text{S}}}$.}
In summary, the proposed {TRICE} framework is given by Algorithm~\ref{Prop}, where in {Step \ref{step10}}, an estimate of ${\bm{H}}_{\text{T}} $ and ${\bm{H}}_{\text{R}} $, {up to trivial scaling factors}, can be obtained from $\widehat{\bm{H}}$ using the LS Khatri-Rao factorization (LSKRF) algorithm proposed in \cite{andre,relay}. 
Please note that Algorithm \ref{Prop} is very general in the sense that any other efficient 2D parameter estimation method can be readily used in Steps 2 and 6, e.g., the methods proposed in \cite{L21Norm,ardah_icassp19,NOMP,URA}.

\begin{algorithm}[t]
{\footnotesize 
	\caption{{\footnotesize {\textbf{T}wo-Stage \textbf{RI}S-aided MIMO \textbf{C}hannel \textbf{E}stimation (TRICE)}}}
	\label{Prop}
	\begin{algorithmic}[1]
		\State{\textbf{Inputs}: Measurement matrix $\bm{Y}$ in (\ref{eq1})}

		\State{\textbf{Stage 1}: Get ${\widehat{\bm{\psi}}}_{\text{T}}$, $\widehat{\bm{\psi}}_{\text{R}}$ using, e.g., OMP or method in \cite{StESBRIT} }

        \State{\textbf{Stage 2}: Assuming knowledge of ${\widehat{\bm{\psi}}}_{\text{T}}$ and $\widehat{\bm{\psi}}_{\text{R}}$ \textbf{do}}

		\State{Get $\underline{\bm{Y}}^{T} = [\underline{\bm{y}}_{1}, \dots, \underline{\bm{y}}_{L}] \in \mathbb{C}^{M_{\text{S}}\times  L } $ from (\ref{eq4})}
		\For{$n = 1$ to $L$}
		
		\State{Get $\hat{\mu}^{\text{h}}_{n}$ and  $\hat{\mu}^{\text{v}}_{n}$ using, e.g., OMP or method in \cite{StESBRIT}}
     	\State{Get $n$th diagonal entry of $\widehat{\bm{G}}$, i.e., $\hat{\alpha}_{n}$ using (\ref{alphn}) }
		\EndFor
        
		\State{Construct $\widehat{\bm{H}} = ( \widehat{\bm{A}}_{\text{T}} \otimes  \widehat{\bm{A}}_{\text{R}}) \widehat{\bm{G}} \widehat{\bm{B}}$ (according to (\ref{h}))}
		\State{Estimate $\widehat{\bm{H}}_{\text{T}} $ and $ \widehat{\bm{H}}_{\text{R}}$ from $\widehat{\bm{H}}$ using \cite[Algorithm 1]{andre} } \label{step10}
	\end{algorithmic}}

\end{algorithm}

\begin{table*}
	\centering
	\caption{\ka{Training overhead and computational complexity analysis}}
	\label{tab}
	\vspace{-5pt}
	\ka{{\scriptsize 
		\begin{tabular}{l|l|l}
			\hline
			Method & Training overhead & Computational complexity \\ \hline
			TRICE-BES & $K_\text{S} \geq L \geq 4$, $N_\text{R} \geq L_\text{R} + 1$, $K_\text{T} \geq L_\text{T} + 1$, $(K_\text{T} - 1)N_\text{R} \geq L$, $(N_\text{R} - 1)K_\text{T} \geq L$ & $\mathcal{O}\big( (N_{\text{R}} K_{\text{T}})^2 K_{\text{S}} + K^3_{\text{S}} + 3L^3 + L\big)$ \\
			TRICE-CS & $N_\text{R} K_\text{T} \approx \mathcal{O}(L\log(\bar{L}_\text{R} \bar{L}_\text{T} /L))$, $K_\text{S} \approx \mathcal{O}(\log(\bar{L}^{\text{v}}_\text{S} \bar{L}^{\text{h}}_\text{S}))$ & $\mathcal{O}(L(N_\text{R} K_\text{T}(\bar{L}_\text{T} \bar{L}_\text{R} + L + L^2)) + 2L^3 + LK_\text{S} \bar{L}^{\text{v}}_\text{S} \bar{L}^{\text{h}}_\text{S}) $ \\ 
			Joint-CS \cite{CSGrid} & $N_\text{R} K_\text{T} K_\text{S} \approx \mathcal{O}(L\log(\bar{L}^{\text{v}}_\text{S} \bar{L}^{\text{h}}_\text{S} \bar{L}_\text{R} \bar{L}_\text{T} /L))$ & $\mathcal{O}(L(N_\text{R} K_\text{T} K_\text{S}(\bar{L}^\text{v}_\text{S} \bar{L}^\text{h}_\text{S}  \bar{L}_\text{T} \bar{L}_\text{R} + L + L^2)) + L^3)$ \\ \hline
	\end{tabular}}}
	\vspace{-5pt}
\end{table*}

\vspace{-20pt}
\section{Numerical Results}	\label{secResults}
\ka{In this section, we show simulation results assuming that the TRICE framework employs, \kr{at} both stages, $(i)$ the 2D DFT-beamspace ESPRIT method from \cite{StESBRIT}, denoted as TRICE-BES, and ($ii$) the on-grid CS method, denoted as TRICE-CS. For comparison, we \kr{also} included the simulation results of the on-grid CS method proposed in \cite{CSGrid}, denoted as Joint-CS. For the CS-based methods, the estimation is {performed} using the classical OMP technique \cite{OMPComp}. To comply with the DFT-beamspace ESPRIT method requirements as discussed in \cite[Lemma 1]{StESBRIT}, we assume that $\psi_{\text{R},\ell} \sim \mathcal{U}(0,{2\pi (N_\text{R} - 1) }/{M_\text{R}})$, $\psi_{\text{T},\ell}\sim \mathcal{U}(0,{2\pi (K_\text{T} - 1) }/{M_\text{T}})$, $\mu^{\text{h}}_{\ell}\sim \mathcal{U}(0,{2\pi (K^{\text{h}}_\text{S} - 1) }/{M^{\text{h}}_\text{S}})$, $\mu^{\text{v}}_{\ell}\sim \mathcal{U}(0,{2\pi (K^{\text{v}}_\text{S} - 1) }/{M^{\text{v}}_\text{S}})$, and the training matrices are chosen as 
$\bm{W}^{T} = [\bm{U}_{M_{\text{R}}}]_{[1:N_\text{R},:]}$,
$\bm{F}^{T} = [\bm{U}_{M_{\text{T}}}]_{[1:K_\text{T},:]} $,
$\bm{Q}^{T}_{\text{h}} = [\bm{U}_{M^{\text{h}}_{\text{S}}}]_{[1:K^{\text{h}}_\text{S},:]} $, and
$\bm{Q}^{T}_{\text{v}} = [\bm{U}_{M^{\text{v}}_{\text{S}}}]_{[1:K^{\text{v}}_\text{S},:]} $, where $\bm{U}_{M}$ denotes the normalized $M\times M$ DFT-matrix. Moreover, we assume that $\alpha_{\ell} \sim \mathcal{CN}(0,1) $ and define {the} $\text{SNR} = \mathbb{E} \big\{ {\Vert \bm{Y} - \bm{Z}\Vert^{2}_{\text{F}}  }/{\Vert \bm{Z}\Vert^{2}_{\text{F}}} \big\}$ and {the} $\text{NMSE} = \mathbb{E} [\Vert \bm{H} - \widehat{\bm{H}} \Vert^2_\text{F}/\Vert \bm{H} \Vert^2_\text{F}]$. Table~\ref{tab} summarizes the training overhead and the complexity of the simulated algorithms \cite{golub13}.
Note that the major difference between TRICE-CS and Joint-CS is that the former decouples the channel parameter estimation into two stages, while the latter jointly estimates them. Therefore, TRICE-CS requires a 2D dictionary in every stage, while Joint-CS requires a single 4D dictionary.}

\ka{In Figs. \ref{Fig1}, \ref{Fig2}, and \ref{Fig3}, we assume that $M_\text{T} = 64, M_\text{R} = 32$, and $M_\text{S} = 256 \text{ } [16\times 16]$. The 4D dictionary for Joint-CS is formed {by} using $64\times 32\times 16 \times 16$ grid points, i.e., it has 524,288 atoms. \kr{On the other hand}, for TRICE-CS, the first stage 2D dictionary is formed {by} using $\beta_{\text{T}} M_{\text{T}} \times \beta_{\text{R}} M_{\text{R}} $ grid points ($\bar{L}_\text{T} = \beta_{\text{T}} M_{\text{T}}$, $\bar{L}_\text{R} = \beta_{\text{R}} M_{\text{R}}$), while the second stage 2D dictionary is formed {by} using $\beta^{\text{v}}_{\text{S}} M^{\text{v}}_{\text{S}} \times \beta^{\text{h}}_{\text{S}} M^{\text{h}}_{\text{S}} $ grid points ($\bar{L}^{\text{v}}_\text{S} = \beta^{\text{v}}_{\text{S}} M^{\text{v}}_{\text{S}}$, $\bar{L}^{\text{h}}_\text{S} = \beta^{\text{h}}_{\text{S}} M^{\text{h}}_{\text{S}}$), where $\{\beta_{\text{T}},\beta_{\text{R}},\beta^{\text{v}}_{\text{S}},\beta^{\text{h}}_{\text{S}}\} \in \{1,2,\dots\}$.}
\begin{figure}
	\centering
	\resizebox{\plotsize}{!}{\begin{tikzpicture}[spy using outlines={rectangle,rounded corners, magnification=4,connect spies,width=5cm,height=1cm}]
\begin{axis}
[
width=\plotwidth,
height=\plotheight,
ymode=log,
xmin=2,
xmax=5,
ymin=1e-5,
ymax= 0,
ylabel={NMSE},
xlabel={SNR [dB]},
xtick={1,2,3,4,5,6,7},
xticklabels = {
	\strut -5,
	\strut 0,
	\strut 5,
	\strut 10,
	\strut 15,
	\strut 20,
},       
% title = $K$ \text{=} 32 ,
%grid=major,
%minor tick num=2,
line width=1.1pt,
mark repeat = 1,
legend entries={LS,TRICE-BES, TRICE-CS (C.1), TRICE-CS (C.2), Joint-CS \cite{CSGrid}},
legend style={
	legend cell align=left,
	legend pos=south west,
	font=\small},
]

%\addlegendimage{only marks, cyan, mark = o}
%\addlegendimage{only marks, blue, mark = square}
%%\addlegendimage{only marks, red, mark = o}    
%%\addlegendimage{only marks, black, mark= triangle}
%%% \addlegendimage{only marks, black,  mark= square}
%%%
%%%
%\addlegendimage{solid, black, no marks}
%\addlegendimage{dashed, black, no marks}
%%%%\addlegendimage{dash dot, black, no marks}

\pgfplotstableread[col sep=comma]{texresults/texFigs/data/fig1_2_2.txt}\tableDataA
%\pgfplotstableread[col sep=comma]{texresults/texFigs/data/fig1_2_3.txt}\tableDataB

\addplot [solid, mark = star,mark options={solid}, draw= black] 
table[x = Ind,  y=d5] from \tableDataA; 
\addplot [solid, mark = triangle,mark options={solid}, draw= red] 
table[x = Ind,  y=d1] from \tableDataA; 
\addplot [solid, mark = square,mark options={solid}, draw= blue] 
table[x = Ind,  y=d4] from \tableDataA; 
\addplot [dashed, mark = square,mark options={solid}, draw= blue] 
table[x = Ind,  y=d2] from \tableDataA; 
\addplot [solid, mark = o,mark options={solid}, draw= cyan] 
table[x = Ind,  y=d3] from \tableDataA;

%\addplot [dashed, mark = triangle,mark options={solid}, draw= red] 
%table[x = Ind,  y=d1] from \tableDataB; 
%\addplot [dashed, mark = square,mark options={solid}, draw= blue] 
%table[x = Ind,  y=d2] from \tableDataB; 
%\addplot [dashed, mark = o,mark options={solid}, draw= black] 
%table[x = Ind,  y=d3] from \tableDataB; 
%\addplot [dashed, mark = square,mark options={solid}, draw= blue] 
%table[x = Ind,  y=d4] from \tableDataB; 

%
%\draw (3.5,0.065) ellipse (0.1cm and 0.1cm);
%\node[text width=3cm] at (4,0.41) {C.1};
%\draw [black, -> ] (3.5,0.075) -- (3.5,0.32);
%
%%\draw (4.5,0.024) ellipse (0.2cm and 0.2cm);
%%\node[text width=3cm] at (5.5,0.003) {C.2};
%%\draw [green, -> ] (4.5,0.016) -- (4.5,0.0042);
%
%
%\draw (3.29,0.02) ellipse (0.1cm and 0.1cm);
%\draw [black, -> ] (3.29,0.017) -- (3.29,0.0051);
%\node[text width=3cm] at (3.8,0.0035) {C.2};

\node[text width=3cm] at (4.2,0.00005) {$L_\text{T} = 2$, $L_\text{R} = 2$};

\node[text width=5cm] at (3.95,3.9) {C.1: $\beta_{\text{T}} =\beta_{\text{R}} =\beta^{\text{v}}_{\text{S}} =\beta^{\text{h}}_{\text{S}} = 1$};
\node[text width=5cm] at (3.95,1) {C.2: $\beta_{\text{T}} =2, \beta_{\text{R}} =4, \beta^{\text{v}}_{\text{S}} =\beta^{\text{h}}_{\text{S}} = 8$};

\end{axis}
\end{tikzpicture}}
	\caption{NMSE vs. SNR. $N_\text{R} = 8$, $K_\text{T} = 8$, and $K_\text{S} = 16 \text{ } [4\times 4]$}
	\label{Fig1}
%	\vspace{-5pt}
\end{figure}
\begin{figure}
	\centering
	\resizebox{\plotsize}{!}{\begin{tikzpicture}[spy using outlines={rectangle,rounded corners, magnification=4,connect spies,width=5cm,height=1cm}]
\begin{axis}
[
width=\plotwidth,
height=\plotheight,
ymode=log,
xmin=2,
xmax=5,
ymin=1e-5,
ymax= 0,
ylabel={NMSE},
xlabel={SNR [dB]},
xtick={1,2,3,4,5,6},
xticklabels = {
	\strut -5,
	\strut 0,
	\strut 5,
	\strut 10,
	\strut 15,
	\strut 20,
},       
% title = $K$ \text{=} 32 ,
%grid=major,
%minor tick num=2,
line width=1.1pt,
mark repeat = 1,
legend entries={TRICE-BES, TRICE-CS (C.2), Joint-CS \cite{CSGrid}, $L = 2 \text{ }(L_\text{T} = 1$\text{, }$L_\text{R} = 2)$, $L = 4 \text{ }(L_\text{T} = 2$\text{, }$L_\text{R} = 2)$, $L = 6 \text{ } (L_\text{T} = 2$\text{, }$L_\text{R} = 3)$ }, 
legend style={
	legend cell align=left,
	legend pos=south west,
	font=\small},
]

\addlegendimage{only marks, red, mark = triangle}
\addlegendimage{only marks, blue, mark = square}
\addlegendimage{only marks, cyan, mark = o}    
%%\addlegendimage{only marks, black, mark= triangle}
%%% \addlegendimage{only marks, black,  mark= square}
%%%
%%%
\addlegendimage{solid, black, no marks}
\addlegendimage{dashed, black, no marks}
\addlegendimage{dash dot, black, no marks}

\pgfplotstableread[col sep=comma]{texresults/texFigs/data/fig1_1_2.txt}\tableDataA
\pgfplotstableread[col sep=comma]{texresults/texFigs/data/fig1_2_2.txt}\tableDataB
\pgfplotstableread[col sep=comma]{texresults/texFigs/data/fig1_2_3.txt}\tableDataC

%\pgfplotstableread[col sep=comma]{texFigs/data/CRB_2_3_6_6_36.txt}\tableDataB

\addplot [solid, mark = triangle,mark options={solid}, draw= red] 
table[x = Ind,  y=d1] from \tableDataA; 
\addplot [solid, mark = square,mark options={solid}, draw= blue] 
table[x = Ind,  y=d2] from \tableDataA; 
\addplot [solid, mark = o,mark options={solid}, draw= cyan] 
table[x = Ind,  y=d3] from \tableDataA;

\addplot [dashed, mark = triangle,mark options={solid}, draw= red] 
table[x = Ind,  y=d1] from \tableDataB; 
\addplot [dashed, mark = square,mark options={solid}, draw= blue] 
table[x = Ind,  y=d2] from \tableDataB; 
\addplot [dashed, mark = o,mark options={solid}, draw= cyan] 
table[x = Ind,  y=d3] from \tableDataB;

\addplot [dash dot, mark = triangle,mark options={solid}, draw= red] 
table[x = Ind,  y=d1] from \tableDataC; 
\addplot [dash dot, mark = square,mark options={solid}, draw= blue] 
table[x = Ind,  y=d2] from \tableDataC; 
\addplot [dash dot, mark = o,mark options={solid}, draw= cyan] 
table[x = Ind,  y=d3] from \tableDataC;

%\addplot [dash dot, mark = triangle,mark options={solid}, draw= red] 
%table[x = Ind,  y=d2] from \tableDataC; 
%\addplot [dash dot, mark = square,mark options={solid}, draw= blue] 
%table[x = Ind,  y=d3] from \tableDataC; 
%\addplot [dash dot, mark = o,mark options={solid}, draw= black] 
%table[x = Ind,  y=d4] from \tableDataC; 

%\addplot [dashed, mark = o,mark options={solid}, draw= cyan] 
%table[x = Ind,  y=d1] from \tableDataB; 
%\addplot [dashed, mark = square,mark options={solid}, draw= blue] 
%table[x = Ind,  y=c1] from \tableDataB; 

%\draw (3.5,0.18) ellipse (0.2cm and 0.3cm);
%\node[text width=3cm] at (5,1.01) {Codebook 1};
%
%\draw (4.3,-22.8) ellipse (0.2cm and 0.7cm);
%

%\node[text width=3cm] at (4.7,-28) {$N_r = 8$};

\node[text width=5cm] at (3.8,1) {C.2: $\beta_{\text{T}} =2, \beta_{\text{R}} =4, \beta^{\text{v}}_{\text{S}} =\beta^{\text{h}}_{\text{S}} = 8$};

\end{axis}
\end{tikzpicture}}
	\caption{NMSE vs. SNR. $N_\text{R} = 8$, $K_\text{T} = 8$, and $K_\text{S} = 16 \text{ } [4\times 4]$}
	\label{Fig2}
%	\vspace{-5pt}
\end{figure}
\begin{figure}[th!]
	\centering
	\resizebox{\plotsize}{!}{\begin{tikzpicture}[spy using outlines={rectangle,rounded corners, magnification=4,connect spies,width=5cm,height=1cm}]
\begin{axis}
[
width=\plotwidth,
height=\plotheight,
ymode=log,
xmin=1,
xmax=5,
ymin=1e-2,
ymax= 0,
ylabel={NMSE},
xlabel={$K_{\text{T}}$},
xtick={1,2,3,4,5},
xticklabels = {
	\strut 4,
	\strut 6,
	\strut 8,
	\strut 10,
	\strut 12,
},       
% title = $K$ \text{=} 32 ,
%grid=major,
%minor tick num=2,
line width=1.1pt,
mark repeat = 1,
legend entries={TRICE-BES, TRICE-CS (C.2), Joint-CS \cite{CSGrid}, $K_{\text{S}} = 4 \text{ } [2\times 2]$, $K_{\text{S}} = 16 \text{ } [4 \times 4]$},
legend style={
	legend cell align=left,
	legend pos=north east,
	font=\small},
]

\addlegendimage{only marks, red, mark = triangle}
\addlegendimage{only marks, blue, mark = square}
\addlegendimage{only marks, cyan, mark = o}    
%\addlegendimage{only marks, black, mark= triangle}
%% \addlegendimage{only marks, black,  mark= square}

\addlegendimage{solid, black, no marks}
\addlegendimage{dashed, black, no marks}
%\addlegendimage{dash dot, black, no marks}

\pgfplotstableread[col sep=comma]{texresults/texFigs/data/fig3_4.txt}\tableDataA
\pgfplotstableread[col sep=comma]{texresults/texFigs/data/fig3_16.txt}\tableDataB
%\pgfplotstableread[col sep=comma]{texresults/texFigs/data/fig3_16.txt}\tableDataC

\addplot [solid, mark = triangle,mark options={solid}, draw= red] 
table[x = Ind,  y=d1] from \tableDataA; 
\addplot [solid, mark = square,mark options={solid}, draw= blue] 
table[x = Ind,  y=d2] from \tableDataA; 
\addplot [solid, mark = o,mark options={solid}, draw= cyan] 
table[x = Ind,  y=d3] from \tableDataA;

\addplot [dashed, mark = triangle,mark options={solid}, draw= red] 
table[x = Ind,  y=d1] from \tableDataB; 
\addplot [dashed, mark = square,mark options={solid}, draw= blue] 
table[x = Ind,  y=d2] from \tableDataB; 
\addplot [dashed, mark = o,mark options={solid}, draw= cyan] 
table[x = Ind,  y=d3] from \tableDataB;

\node[text width=3cm] at (2.2,0.025) {$L_\text{T} = 2$, $L_\text{R} = 2$};

\node[text width=5cm] at (2.75,0.015) {C.2: $\beta_{\text{T}} =2, \beta_{\text{R}} =4, \beta^{\text{v}}_{\text{S}} =\beta^{\text{h}}_{\text{S}} = 8$};

%
%
%\addplot [dash dot, mark = triangle,mark options={solid}, draw= red] 
%table[x = Ind,  y=d1] from \tableDataC; 
%\addplot [dash dot, mark = square,mark options={solid}, draw= blue] 
%table[x = Ind,  y=d2] from \tableDataC; 
%\addplot [dash dot, mark = o,mark options={solid}, draw= black] 
%table[x = Ind,  y=d3] from \tableDataC; 

%\addplot [dashed, mark = triangle,mark options={solid}, draw= red] 
%table[x = Ind,  y=d1] from \tableDataB; 
%\addplot [dashed, mark = square,mark options={solid}, draw= blue] 
%table[x = Ind,  y=d2] from \tableDataB; 
%\addplot [dashed, mark = o,mark options={solid}, draw= black] 
%table[x = Ind,  y=d3] from \tableDataB; 
%\addplot [dashed, mark = square,mark options={solid}, draw= blue] 
%table[x = Ind,  y=d4] from \tableDataB; 

%\draw (3.5,0.085) ellipse (0.1cm and 0.1cm);
%\node[text width=3cm] at (4.05,0.21) {C.1};
%\draw [green, -> ] (3.5,0.095) -- (3.5,0.18);
%
%%\draw (4.5,0.024) ellipse (0.2cm and 0.2cm);
%%\node[text width=3cm] at (5.5,0.003) {C.2};
%%\draw [green, -> ] (4.5,0.016) -- (4.5,0.0042);
%
%
%\draw (3.5,0.0135) ellipse (0.1cm and 0.1cm);
%\node[text width=3cm] at (4.5,0.02) {C.2};
%\draw [green, -> ] (3.55,0.014) -- (3.85,0.019);

\end{axis}
\end{tikzpicture}}
	\vspace{-10pt}
	\caption{NMSE vs. $K_\text{T}$ and $K_\text{S}$. SNR = 5 dB.}
	\label{Fig3}
	\vspace{-10pt}
\end{figure}
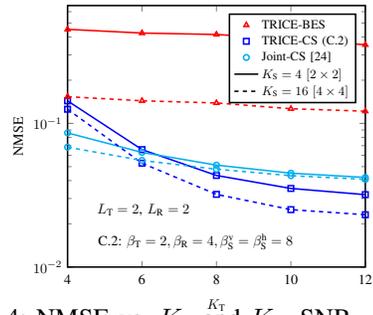

\ka{From Fig. \ref{Fig1}, \kr{in case of C.1}, we can see that TRICE-CS approaches the Joint-CS performance as the SNR increases, since in this case both methods have the same grid resolution, while Joint-CS outperforms TRICE-CS in the low SNR regime, due to its joint estimation. However, by increasing the grid resolutions as in C.2, TRICE-CS outperforms Joint-CS even in the low SNR regime. Note that, \kr{using the C.2 case,} TRICE-CS has a much lower complexity when compared to Joint-CS, since it has $\approx94\%$ less atoms. \kr{By its turn}, TRICE-BES has a good performance in the medium and the high SNR regimes, where Fig.~\ref{Fig2} shows that TRICE-BES provides a satisfactory performance in case of very sparse channels, \kr{cf. the $L = 2$ case}. Further, Fig.~\ref{Fig3} shows that the estimation accuracy can be improved by increasing $K_\text{T}$ and/or $K_\text{S}$. }

\section{Conclusions}
%\vspace{-5pt}
{The proposed TRICE framework} is a two-stage channel parameter estimation scheme for single-user \gls{ris}-aided MIMO mmWave systems. By exploiting the low-rank nature of mmWave channels and by decoupling the channel parameter estimation problem into two stages, we have shown that {TRICE} not only {has} a high estimation performance, but also {affords} a low training overhead and {has} a low computational complexity, which makes it appealing in practical applications.

\newpage	
\bibliographystyle{IEEEtran}
\bibliography{refs}		
\end{document}